\documentclass[aps,pra,amsmath,amssymb,twocolumn,groupedaddress]{revtex4-1}
\newcommand{\be}{\begin{equation}}
\newcommand{\ee}{\end{equation}}
\newcommand{\ba}{\begin{eqnarray}}
\newcommand{\ea}{\end{eqnarray}}
\newcommand{\baa}{\begin{eqnarray}}
\newcommand{\eaa}{\end{eqnarray}}
\newcommand{\ed}{\end{document}}
\newcommand{\lab}[1]{\label{#1}}
\newcommand{\re}[1]{(\ref{#1})}
\newcommand{\ci}[1]{\cite{#1}}
\def\pa{\partial}

\usepackage{graphicx}
\usepackage{color}

\usepackage{multirow}

\begin{document}

\title{High harmonic generation  under dynamical confinement: An atom-in-box model}

\author{S. Rakhmanov$^a$, D. Matrasulov$^b$, B. Eshchanov$^a$ and V.I. Matveev$^{c,d}$}

\address{$^a$ Physics Department, National Universty of Uzbekistan, Vuzgorodok, 100174,  Tashkent,
Uzbekistan\\
$^b$ Turin Polytechnic University in Tashkent, 17 Niyazov Str.,
100095,  Tashkent, Uzbekistan\\
$^c$ Federal Center for Integrated Arctic Research, Ural Branch,
Russian Academy of Sciences, nab. Severnoi Dviny 23, Arkhangelsk,
163000 Russia\\
$^d$ Northern (Arctic) Federal University, nab. Severnoi Dviny 17,
Arkhangelsk, 163002 Russia}


\vspace{10pt}

\begin{abstract}
We consider optical high harmonic generation in a hydrogen atom
confined in a breathing spherical box by considering atomic
nucleus as fixed at the center of sphere. In such spherically
symmetric, dynamical trap, the high harmonic generation spectrum
is calculated at different values of the oscillation  amplitude,
confinement size and atomic nucleus charge.
\end{abstract}

%
%
%
%
%
\maketitle

\section{Introduction}

Study of the nonlinear optical phenomena in the interaction of
atoms and molecules with external optical fields is of practical
and fundamental importance for the problems, e.g., attosecond
physics, high-power laser generation, optical materials design and
optoelectronic device fabrication, etc. An interesting aspect of
this topic is optical harmonic generation in quantum regime, which
attracted much attention recently \ci{Tong}-\ci{Boydbook}.  One of
the main tasks in this field is achieving  non-, or
slowly-decaying intensity of the  generated high harmonics.
Usually, in free atoms and molecules interacting with external
optical fields intensity of generated harmonic  decreases, as the
order of harmonic increases. Such effect makes difficult
generating of very high order harmonics and ultrashort pulses, as
their intensity becomes very small. Therefore revealing the
regimes for high harmonic generation, providing  non-decaying
intensity (as a function of harmonic order) is of importance for
different practically tasks. One of such ways could be confining
of atoms and molecules in finite spatial domain, where by changing
confinement size and geometry one can   control the harmonic
generation.

In this paper we discuss a model for high harmonic generation in
an atom confined in a dynamical (time-dependent) trap. The latter
presents a spherical box with harmonically breathing radius. The
nucleus of atom is considered as fixed at the center of sphere, so
that spherical symmetry of whole system is not broken.It is well
known that confined systems have completely different physical
properties compared to the free ones. Such difference is caused by
the boundary conditions imposed for the Schrodinger equation,
which cause modification of the energy and wave function spectrum.

                                                                                                                                                                                                                                                                                                                                                                                                                                                                                                                                                                                                                                                                                                                                                                                                                                                                                                                                                                                                                                                                                                                                                                                                                                                                                                                                                                                                                                                                                                                                                                                                                                                                                                                                                                                                                                                                                                                                                                                                                             Atoms and molecules confined in nanoscale domains have the
The earlier studies of atom-in-box system  date back to the Refs.
\ci{Michels,Sommerfeld}, where effect of the pressure on an atom
was explored in quantum approach. Later, Wigner studied the
problem  within the Rayleigh-Schr\"odinger perturbation theory and
showed that in the limit of infinite box size, the result does not
converge into that for the free atom. Considerable number of
papers on the atom-in-box problem (see, e.g.,
\ci{Weil}-\ci{Burrows03} and review paper \ci{Jaskolski} for more
references) has been published, since from these pioneering works.
In \ci{Weil} the problem of hyperfine splitting in such system is
treated, Ref.\ci{Rubinstein} presents first numerical solution of
the problem.  More comprehensive treatment of atom-in-box system
can be found in a series of papers by Burrows et.al
\ci{Burrows0,Burrows,Burrows03,Burrows04}, where the authors used
different analytical and numerical methods for finding eigenvalues
of the system. In \ci{Dragoslav,Lumb,Dragoslav1}  the quantum
dynamics of hydrogen atom confined in a spherical box and driven
by external electric field is studied.

Experimentally, atom-in-box system can be realized, e.g., in
co-called atom optic billiards which represent a rapidly scanning
and tightly focused laser beam creating a time-averaged
quasi-static potential for atoms  \ci{Raizen}-\ci{Douglas}. By
controlling the deflection angles of the laser beam, one can
create various box (billiard) shapes. Another method  is putting
the hydrogen atom inside the fullerene \ci{Connerade1,Connerade2}.
Different ways for confining of atom inside a cages in experiment
are discussed in \ci{Jaskolski}. We note that all the studies of
atom-in-box system are mainly focused by considering the case of
box static boundaries, while modern technologies in quantum and
atom optics provide different tools for creating dynamical
confinement.  Some versions for experimental realization of such
dynamical traps have already been discussed in the literature (
see, e.g.,the Refs.\ci{TDT1,TDT2,TDT3,TDT4}). In such cases the
dynamics of atomic electron is completely different than that for fixed boundaries.\\
In this paper we consider high harmonic generation in quantum
regime by an atom confined in a "breathing" spherical box. To
solve the time-dependent Schrodinger equation for Coulomb
potential with dynamical boundary conditions we use the same
approach as that in our recent paper \ci{DMEPJD}.

We note that the problem of moving boundaries in quantum mechanics
is treated in terms of the  Schr\"{o}dinger equation with
time-dependent boundary conditions. Earlier, the quantum dynamics
of a particle confined in a time-dependent box was studied in
different contexts (see Ref. \ci{doescher}-\ci{Our02}). Here we
consider similar problem for an electron moving in a Coulomb field
of the atomic nucleus, confined in a harmonically breathing
spherical box.

This paper is organized as follows. In the next section we give
brief description of a quantum system, consisting of one-electron
atom confined in a breathing spherical box. Section 3 presents
detailed treatment of high harmonic generation in sauch system.
 Finally, section 4 provides some concluding remarks.

\section{Quantum dynamics of  by hydrogen-like atom confined in a time-dependent spherical box}

Consider atom confined in a spherical box with time-varying radius
given by $r_0 = r_0(t)$. In this case the sphere retains its shape
during the expansion(contraction), so that the the central
symmetry is not broken. Therefore, if atomic nucleus is fixed at
the center of sphere, the electron dynamics is described by the
time-dependent radial Schr\"{o}dinger equation which is given as
\be i\frac{\partial R(r,t)}{\partial t}=\hat{H}R(r,t), \lab{TDSE1}
\ee where \baa \hat{H} =-\frac{1}{2}\frac{\partial^2 }{\partial
r^2}-\frac{1}{r}\frac{\partial }{\partial r}+\frac{l(l+1)}{2 r^2}
-\frac{Z}{r}.  \nonumber \eaa For such regime, the boundary
conditions for Eq.\re{TDSE1} are imposed as \be
R(r,t)|_{r=r_0(t)}=0.\lab{bc01} \ee

To solve Eq.\re{TDSE1} one should  reduce  the boundary conditions
into time-independent form. This is can done by using the
following coordinate transformation \ci{mak91,mak92}: \be
 y=\frac{r}{r_0(t)}. \lab{tr1}
\ee

\begin{figure}[t!]
\includegraphics[totalheight=0.23\textheight]{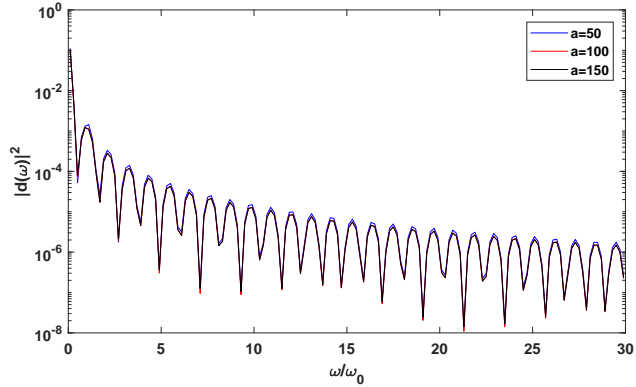}
 \caption{ (Color online) Harmonic generation spectrum for hydrogen atom in a breathing spherical box  at different
 $a$ and for fixed frequency $\omega_0 = 1$, amplitude $b = 10$ and nucleus charge $Z=1$. } \label{fig:1}
\end{figure}

\begin{figure}[t!]
\includegraphics[totalheight=0.23\textheight]{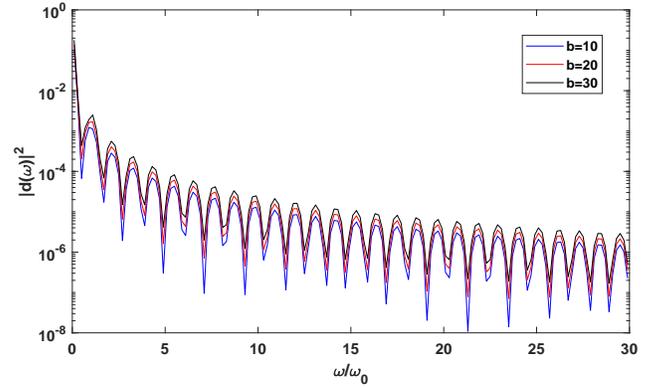}
 \caption{ (Color online)   Harmonic generation spectrum for hydrogen atom in a breathing spherical box  at different
oscillation amplitudes, b and for fixed frequency $\omega_0 = 1$,
$a = 100$ and nuclear charge $Z=1$. } \label{fig:2}
\end{figure}

In terms of new coordinate, $y$ Eq.\re{TDSE1} can be rewritten as
\baa i\frac{\partial R(y,t)}{\partial t}=\Biggl[-\frac{1}{2
r_0^2}\frac{\partial^2 }{\partial y^2}-\Biggl( \frac{1}{2 r_0^2
y}-i\frac{\dot{r}_0}{r_0}y \Biggl)\frac{\partial }{\partial y}
\nonumber \\ +\frac{l(l+1)}{2r_0^2
y^2}-\frac{Z}{r_0y}\Biggl]R(y,t) \equiv \hat{\tilde H} R(y,t).
\lab{TDSE2} \eaa In Eq.\re{TDSE2} the self-adjointness is broken,
i.e., operator $\hat{\tilde H}$ is not Hermitian. In addition, due
to the first-order derivative in this equation makes complicated
its solution. Therefore, to restore Hermitticity and remove the
first order derivative, one can use the transformation of the wave
function which is given by \be
 R(y,t)=\frac{1}{r_0(t)^{3/2} y}e^{\frac{i}{2}r_0(t)\dot{r}_0(t)
 y^2}\Phi(y,t).
\lab{30} \ee  Doing such transformation  and  introducing of the
new time-variable defined as \ci{mak91,razavy,Our01}

\baa \tau =\int_0^{t}\frac{ds}{r_0(s)^2}, \nonumber \eaa

we reduce Eq.\re{TDSE2} into the Hermitian for which can be
written as \be i\frac{\partial \Phi}{\partial
\tau}=-\frac{1}{2}\frac{\partial^2 \Phi}{\partial y^2}+\Biggl(
\frac{1}{2}r_0^3 \ddot{r}_0 y^2+\frac{l(l+1)}{2y^2}-\frac{Zr_0}{y}
\Biggl)\Phi. \lab{TDSE3} \ee The boundary condition for $\Phi$ is
imposed as
\baa \Phi(y,t)|_{y=1} =0. \nonumber \eaa

We note that Eq.\re{TDSE3} can be obtained from Eq.\re{TDSE1}  by
using following unitary transformation for the Hamiltonian $H$
\ci{razavy,Our01}:

\baa \hat{\tilde H} =e^{-iV}e^{-iU}
(\hat{H}-i\frac{\partial}{\partial t})e^{iU}e^{iV}, \nonumber \eaa

where

\baa U=i(r\frac{\pa}{\pa r}+\frac{3}{2})\ln r_0(t), \nonumber \eaa

and

\baa V =-\frac{1}{2}r_0\frac{dr_0}{d\tau}y^2. \nonumber \eaa

Eq.\re{TDSE3} is the Schr\"{o}dinger equation for an electron
moving in the field of Coulomb and time-dependent harmonic
oscillator potentials. The charge of the Coulomb field is
time-dependent due to the factor $r_0(t)$. The whole system is
confined in a spherical box with unit radius. Time and coordinate
variables cannot be separated in Eq.\re{TDSE3} and one needs to
solve it numerically. To do this we expand $R(y,t)$ in terms of
the complete set of eigenfunctions of a spherical box with unit
radius:

\be \Phi(y,t) =\sum_{nl} C_{nl}(t)\varphi_{nl}(y),  \lab{tr22} \ee

where $\varphi_{nl}(y) =N_{nl}y j_l(\lambda_{nl}y)$ are the
eigenfunctions of the stationary Schr\"{o}dinger equation for a
spherical box  of unit radius, $j_l$ are the spherical Bessel
functions.  Inserting this expansion into Eq.\re{TDSE3} we get the
system of first order differential equations with respect to
$C_{nl}(t)$: \be i\dot {C}_{nl}(t) =r_0^{-2}\sum_{n'l'}
C_{n'l'}(t) V_{nln'l'}(t) +\varepsilon_{nl}r_0^{-2} C_{nl},
\lab{system1} \ee where $\varepsilon_{nl}$ are the eigenvalues of
the Schrodinger equation for spherical box and

\baa V_{nln'l'}(t) =<\varphi_{n'l'}|-\frac{Zr_0}{y}+\frac{1}{2}r_0^3
\ddot{r}_0 y^2|\varphi_{nl}>. \nonumber \eaa

In solving Eqs.\re{system1} numerically one should take into account
the normalization condition for the expansion coefficients,
$C_{nl}$:

\baa \sum_{nl} |C_{nl}(t)|^2 =1, \nonumber \eaa

which follows from the normalization condition for the wave
function:

\baa \int_0^{r_0(t)}|\Psi(r,t)|^2 r^2dr =1. \nonumber \eaa

Initial conditions for Eq.\re{system1} are imposed as an atomic
state with a given set of quantum numbers, $(n_1,l_1)$, i.e., all
coefficients, $C_{nl}=0$, except $C_{n_1l_1}$.

Having found the wave function $\Psi(r,t)$,  one can compute the
main characteristics of high harmonic generation, i.e., the
average dipole moment, which is given by
$$
\bar d(t) =-<R(r,t)|r|R(r,t)>,
$$
where the wave function $R(r,t)$ is related to $\Phi(y,t)$ via
Eq.\re{30}.\\ We are interested in the study of optical harmonic
generation in the system described by Eqs.\re{TDSE1} and
\re{bc01}. The spectrum of harmonic generation is characterized by
the power spectrum, i.e,, absolute square of the Fourier transform
of the average dipole moment, which is given by \ci{Tong}

\be |\bar{d}(\omega)|^2=|\frac{1}{T}\int_0^T e^{-i\omega
t}\bar{d}(t)dt|^2,\lab{ps1}, \ee where $T$ is the interaction
time.

\begin{figure}[t!]
\includegraphics[totalheight=0.23\textheight]{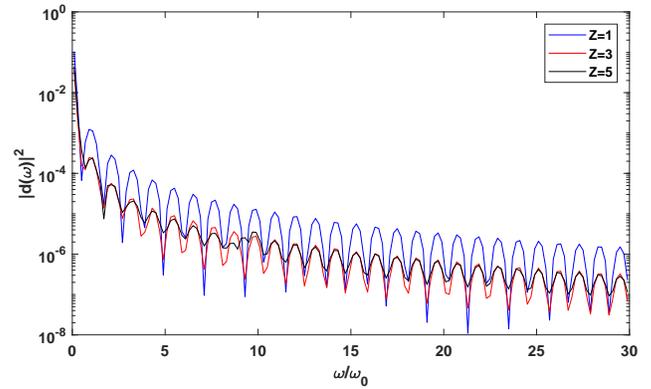}
 \caption{ (Color online)  Harmonic generation spectrum for hydrogen atom in a breathing spherical box  at different
values of the atomic nuxcleus charge,  $Z$ for fixed fixed values
of the frequency $\omega_0 = 1$, amplitude $b=10$ and $a = 100$. }
\label{fig:4}
\end{figure}

\section{High harmonic generation by hydrogen-like atom confined in a breathing spherical box}

Atom-in=-box system described by Eqs.\re{TDSE1} and \re{bc01} have
been studied in the Ref.\ci{DMEPJD} by focusing on the atomic
electron's dynamics. Here we will consider high harmonic
generation in this system caused by the interaction of atomic
electron with the oscillating wall. In atom optic experiments such
a "wall" can be created e.g., by an optical field.  In the
following we will focus on the regime, when the wall of the box is
harmonically oscillating, i.e., in harmonically breathing regime,
where the radius if given by $r_0(t)= a + b\cos\omega_0 t$, where
$\omega_0$ and $b$ are the oscillating frequency and amplitude,
respectively.

For the "virtual" system described by Eq.\re{TDSE3} the
oscillating potential term can be written in the form of
multichromatic field including 4 different frequencies as

$$V(y,t)=\frac{1}{2}r_0^3\ddot{r}_0y^2=$$

\be \frac{1}{2}(A+B\cos\omega_0 t+C\cos2\omega_0 t +D\cos3\omega_0
t+E\cos4\omega_0 t)y^2 \ee

where

$$ A=-\frac{3b^2\omega_0^2}{8}(4a^2+b^2), \hspace{2mm}
B=-\frac{3ab\omega_0^2}{4}(4a^2+9b^2),$$

$$\hspace{2mm} C=-\frac{b^2\omega_0^2}{2}(3a^2+b^2), \hspace{2mm}
D=-\frac{3ab^3\omega_0^2}{4},  \hspace{2mm}
E=-\frac{b^4\omega_0^2}{8}$$

Thus the virtual system can be considered as a quantum box with
unit size and driven by nonlinearly polarized multichromatic
field.  In Fig. 1 spectra of high harmonic generation are plotted
at different values of the box's initial radii (for $\omega_0 =
1$,  $b = 10$ and $Z=1$). The plots show that harmonic generation
spectrum is not sensitive to the change of the confinement size.
Of course, this does not concern very large box sizes comparable
with the free atom. Fig. 2 presents similar plots  (for $\omega_0
= 1$, $a = 100$ $Z=1$) at the different values of the breathing
amplitude, $b$. In this case, the power spectrum is sensitive to
the wall's oscillation amplitude and as $b$ is higher, as larger
is the harmonic generation intensity. Finally, in Fig. 3 high
harmonic generation spectra are compared for different values of
the atomic nucleus charge, $Z$ (for $\omega_0 = 1$,  $b=10$ and $a
= 100$). For all plots the interaction time is taken as $T=100$.
It is useful to compare the above plots with those from the
Ref.\ci{Tong}, where the high harmonic generation by the free
hydrogen atom interacting with external optical field is studied.
First different, which can be seen at first glance is the
considerable (3 to 5 order) quantitative difference between the
generation spectra. This implies that the role of dynamical
confinement can be crucial in harmonic generation and may provide
effective tool for tuning the process.  Well-known features of the
emission spectra of an atom in a strong laser field, known as
''the plateau'' and ''the cutoff,'' are a wide frequency region
with harmonics of comparable intensities, and an abrupt intensity
decrease at the highenergy end of the plateau. Using the regime of
dynamical confinement, one may achieve the regime, the "cur-off"
frequency can be increased. Possible increasing the cut-off
frequency has been discussed earlier in  \ci{Rost}

\section{Conclusions}
In this work we studied high harmonic generation by hydrogen-like
atom under the dynamical confinement created by impenetrable
spherical box with time-dependent radius. The main focus of the
study is given to the dependence of high harmonic generation
intensity on the box initial size, oscillation amplitude and
atomic nucleus charge. The analysis of  high harmonic generation
spectra is done using the numerical solutions of time-dependent
Schrodinger equation for Coulomb potential, for which the
time-dependent box boundary conditions are imposed. Behavior of of
the Fourier transform (power spectrum) of the average dipole
moment is studied for different box breathing regimes, as well as
for for different atomic nucleus charges. By analyzing the high
harmonic generation spectra in the dynamical confinement regime we
found that the intensity of generation is much higher than that
for unconfined (free) atom. This implies that the above model of
high harmonic generation can be very effective proposal
 for attosecond pulse generation.

\section*{Acknowledgement}
This work is partially supported by the grant of the Ministry of
Innovation Development of Uzbekistan (Ref. No. BF-2-022).

\end{document}